\documentstyle[11pt,newpasp,twoside,epsf]{article}
\def\edcomment#1{\iffalse\marginpar{\raggedright\sl#1\/}\else\relax\fi}
\reversemarginpar

\begin{document}

\title{A trial symbolic dynamics of the planar three-body problem}
\author{Kiyotaka TANIKAWA$^1$ and Seppo MIKKOLA$^2$}

\affil{$^1$National Astronomical Observatory of Japan\\
 Mitaka 181-8588, Japan; tanikawa.ky@nao.ac.jp\\
$^2$Tuorla Observatory, University of Turku, Turku, Finland;\\
seppo.mikkola@utu.fi
}

\begin{abstract}
 
Symbolic dynamics is applied to the planar three-body problem. 
Symbols are defined on the planar orbit when it experiences a  
syzygy crossing. If the body $i$ is in the middle at the  
syzygy crossing and the vectorial area of the triangle made with 
three bodies changes sign from $+$ to $-$, number $i$ is given to 
this event, whereas if the vectorial area changes sign from $-$ to
$+$, number $i+3$ is given. 
We examine the case of free-fall three-body problem for the first
few digits of symbol sequences, and we examine the case 
with angular momentum only for the first digit of the symbol sequences. 
This trial experiments show some new aspects of the planar three-body
problem.     
\end{abstract}


\section{Introduction}

Symbolic dynamics for the planar three-body problem is not yet fully
developed. Not many authors are involved in this direction 
(Chernin et al. 2006; Myllari 2007; Moeckel 2007). 
We need to find a good procedure to assign symbols comparable to 
the one-dimensional case (Tanikawa \& Mikkola 2000). 

In the free-fall three-body problem, we saw that binary collision curves 
(formed with initial condition points of orbits which experience collision) 
constitute the boundary of regions in the initial value plane 
(Tanikawa \&Umehara, 1998). 
We expect this may be the case also for the case with angular momentum.
In the present report, we introduce symbols so that the boundaries of 
different symbol sequences are binary collsion curves. 
After the definition of symbols, we start numerical symbolic dynamics 
of the planar three-body problem with angular momentum extending 
the free-fall problem. 

\section{Introduction of symbols}

We introduce symbols in this section. First, we define the 
signed area of the triangle formed with three bodies. 
If the three bodies are arranged in the counter-clockwise order
for body numbers 1, 2, and 3, we consider that the area is positive 
(see Fig. 1
). 
If the order is reverse, the area is considered to be negative. 
The absolute value of the area of the triangle is its usual area. 

Using this definition of area, we assign a symbol to a particular event 
on the orbits. Suppose that the configuration of the three bodies becomes 
collinear. If the angular momentum of the system is not zero, this 
configuration cannot be maintained for a finite non-zero time interval
except the rectilinear case of the three-body problem. 
Before and after this configuration, triangles of non-zero area are 
recovered. 

\begin{figure}[h] 
\begin{center}
\plotfiddle{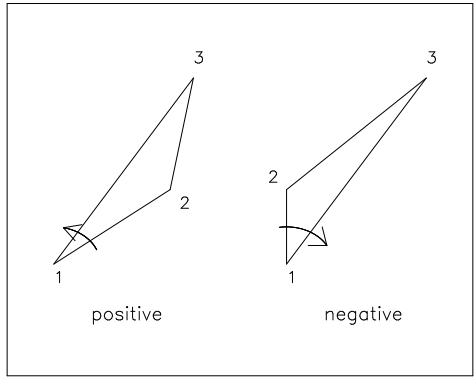}{4 cm}{0}{50}{50}{-180}{-40}
\end{center}
\caption{ Vectorial area of triangles}
\label{vectorial-area}
\end{figure}

We ask here whether the middle particle may cross or only be tangent to 
the syzygy line. The latter does not happen since otherwise 
the trajectory of the middle body is convex to the remaining two
bodies when it is on the syzygy. This is impossible because the trajectory
of any body should be concave to either or both of the remaining bodies 
due to the gravitational attraction. 

We need to consider the case of binary collision. At binary collision, 
the configuration becomes collinear. As is well-known, the trajectory 
of a body relative to the other body is a parabola in the vicinity of 
binary collision. The third body can be considered to be stand still 
compared with the high speed of the binary components. This again shows
that the collinear configuration cannot be maintained before and after 
collision.

Finally we need to make an important remark. At binary collision, 
an orbit experiences syzygy crossings not only once. The orbit may 
experience at most three syzygy crossings. 
In addition, the number of crossings is different depending on 
whether the collision is considered as a limit of retrograde encounter 
or prograde encounter. The limit should be taken to keep the continuity 
to the neighboring initial conditions. 
The analysis of collision will be given elsewhere. 
We here make a notice that two or three symbols may be given 
to the orbit at collision.

Now, suppose that the area changes sign from $+$ to $-$ at some 
instant. Then we give 
symbol '1' if the longest edge is 2--3, that is, the edge connecting
body2 and body 3. Simlarly, we give symbol '2' 
if the longest edge is 3--1, and  we give symbol '3' 
if the longest edge is 1--2. When the area changes sign from $-$ to 
$+$, we give symbol '4' if the longest edge is 2--3.
Simlarly, we give symbol '5' if the longest edge is 3--1, and  
we give symbol '6' if the longest edge is 1--2. 
See Fig. 2
.

\begin{figure}[h] 
\begin{center}
\plotfiddle{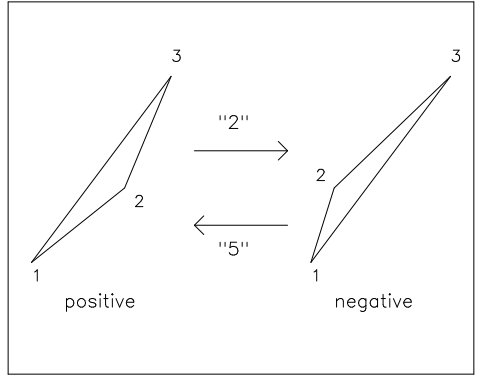}{4 cm}{0}{50}{50}{-180}{-40}
\end{center}
\label{symbolass}
\caption{ Assignment of symbols.}
\end{figure}

\section{Orbits and Symbol sequences}

Each time a triple system becomes collinear, a symol is given according 
to the rule described in the preceding section except at collision. 
Therefore, an orbit represented by a continuous curve in the phase space 
is replaced by a bi-infinite symbol sequence. 
Here we only consider the future symbol sequence. 
We denote the present by a point, and symbols by $s_i$, then a symbol 
sequence $s$ can be written as 
$$
     s = .s_1 s_2 s_2 \ldots .   \eqno(1)
$$ 

\subsection{The planar system with angular momentum}

Let us briefly explain the initial conditions of numerical integrations.
We want to extend the free fall problem with equal masses 
(Anosova \& Orlov 1986; Tanikawa et al. 1995), in which bodies 2 and 3 
initially stand still at $(-0.5,0)$ and $(0.5,0)$, respectively, 
whereas body 1 stands still at 
$(x,y) \in D_{11} = \{ x \geq 0, y \geq 0, (x-0.5)^2 + y^2 \leq 1 \}$. 
$D_{11}$ is called the initial condition plane 
(Fig. 3
).
If the body 1 moves everywhere in $D_{11}$, all the possible triangles are 
realized. 

Now, we give velocities to the bodies and angular momentum to the
system still with equal masses. In the planar three-body problem with 
angular momentum, there too many degrees of freedom. We need somehow 
to restrict the initial condition space so as to be able to express the
numerical results visibly. Here, we give maximal angular momentum for 
a given configuration triangle (See Kuwabara \& Tanikawa, this isuue).
 In this case, one of the symmetries 
is lost, so the initial condition space becomes doubled. 
See the right panel of Fig. 3
. 
This time $D_{11} \cup D_{12}$ is the initial condition plane
(Tanikawa \& Kuwabara 2007).

\begin{figure}[htbp]
\begin{center}
\plotfiddle{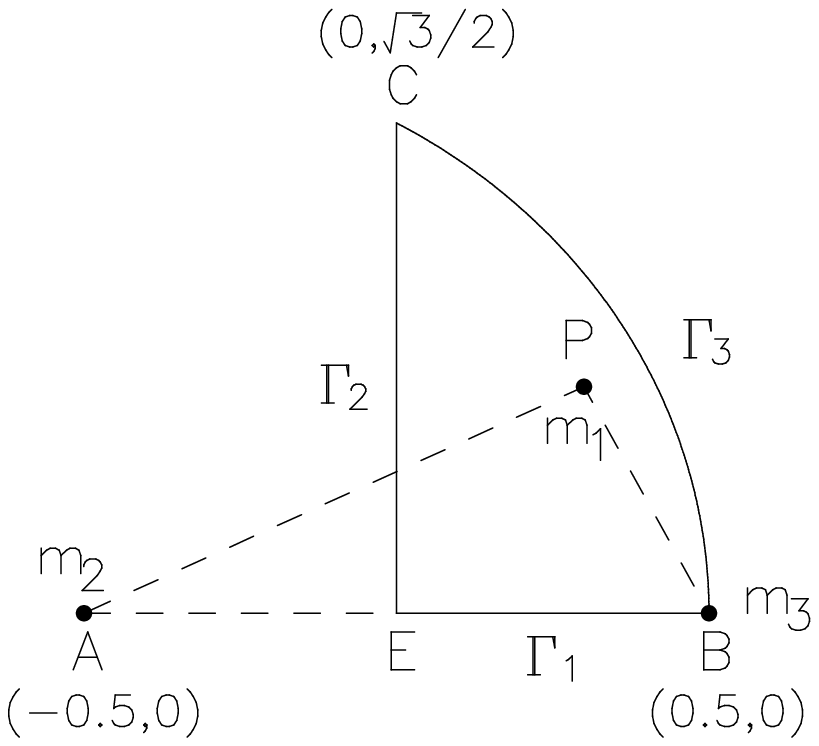}{2 cm}{0}{60}{60}{-300}{-180}
\plotfiddle{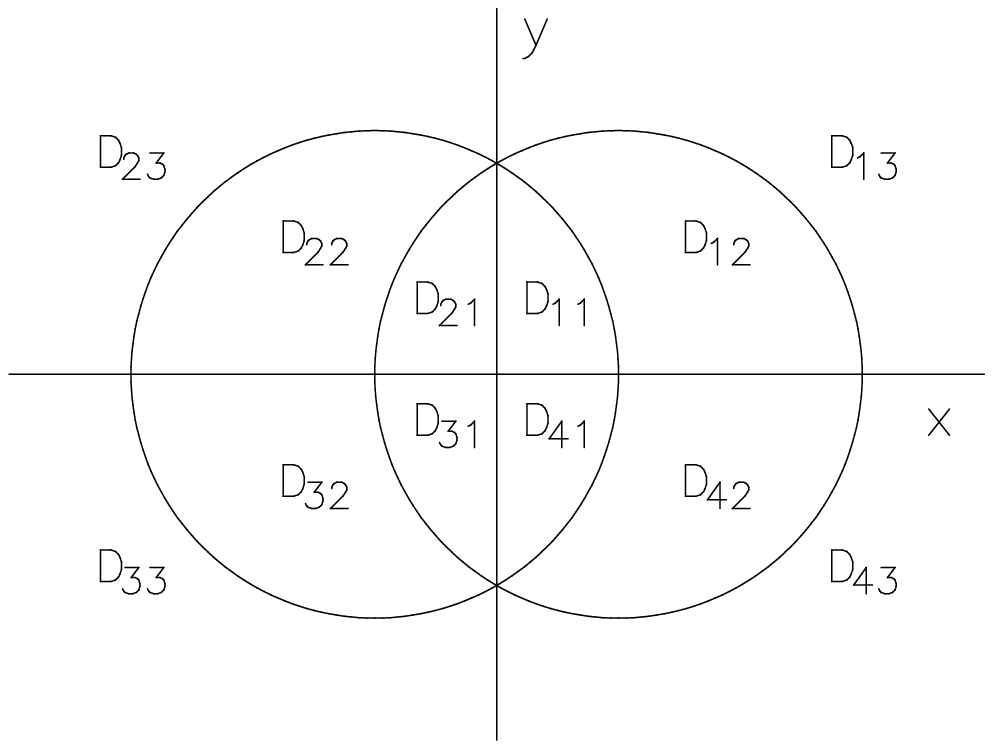}{2 cm}{0}{60}{60}{-80}{-100}
\end{center}
\label{geomerty}
\caption{Geometry of the free-fall initial conditions (left) and 
the problem with angulur momentum (right)}
\end{figure}

\subsection{Boundaries of symbols in the initial condition space}

Now, suppose that all the points in the initial condition space 
have their own symbol sequences (1), that is, the orbits starting 
at points of the initial condition space are all integrated. 
If we truncate symbol sequences at the $n$-th digit, there are 
finite number of possible combinations of symbols in these
length-$n$ ``words''. The initial condition space is divided by 
``cylinders'' which contain these words. 
We ask what kind of points, or equivalently, orbits, constitute 
these boundaries. 

For simplicity of discussion, let us consider the case $n=1$, the first
digit. It is apparaent from Fig. 4 that initial tirangles have positive
areas, so only '1', '2', and '3' appear at the first syzygy crossing. 
These three symbols divide the initial condition space. Suppose the
region where the first digit is '1'. Take orbit $O$ in which body 
1 crosses the syzygy between bodies 2 and 3 and at finite non-zero 
distances both from bodies 2 and 3. 
Then, all neighboring orbits also experience syzygy
crossing of body 1 bewteen bodies 2 and 3. This means that orbit 
$O$ is inside the region occupied by symbol '1' which we call region
'1. This argument 
applies as long as body 1 crosses the syzygy at finite distances from 
both bodies 2 and 3. Orbits of the boundaries of region '1' should be 
collision orbits. There can be two boundaries: between regions '1' and 
'2' and between regions '1' and '3'. Similarly, there can be two
boundaries of region '2', and two boundaries of region '3'. In general,
there are six symbols. So, the boundaries of region '$i$' ($i=1,2,\ldots$)
can be more than two.

\begin{figure}[htbp] 
\begin{center}
\plotfiddle{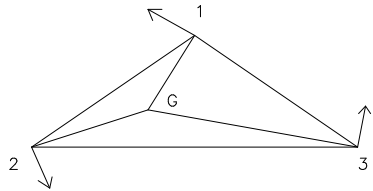}{2 cm}{0}{50}{50}{-150}{-70}
\end{center}
\label{triangle-init}
\caption{ Initial triangle and the momentum}
\end{figure}

\section{Numerical results}

\subsection{The free-fall problem}

In order to justify the argument of \S 3.2, we first give 
results on the free-fall problem (Tanikawa \& Umehara 1998). 
Figure 5 shows how the initial condition plane is divided by 
collision curves. It is to be noted that the boundaries of 
the plane, the $x$-axis, $y$-axis, and the circular boundary 
are all collision curves. 

Our numerical results are shown in Fig.6. In the free-fall case, 
body 1 passes through between bodies 2 and 3 at the first syzygy 
crossing. So the whole space is occupied by region '1'. 
The initial condition space divided by '15' and '16' at the first two 
digits. The reason is simple. At the right part of the space, bodies 
1 and 3 form a binary, so body 3 passes through between bodies 1 and 2, 
and the vectorial area goes from '-' to '+', giving symbol '6'.  
At the right part, body 2 passes through the syzygy, giving symbol '5'.
The boundary of two regions is the $y$-axis. This is seen in the figure 
of the left panel of Fig. 6. In the lower-left and lower-right corners 
of the figure, some different colors are seen. These are considered to
be numerical artifacts.

\begin{figure}[htbp] 
\begin{center}
\plotfiddle{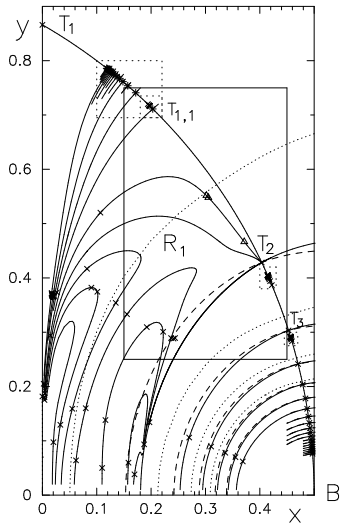}{6 cm}{0}{40}{40}{-140}{-50}
\end{center}
\label{free-fall}
\caption{ binary collision curves divide the plane
(Reproduced from Tanikawa \& Umehara (1998)).
}
\end{figure}

In the middle panel of Fig. 3, the initial plane is divided by 
regions with the first three digits of symbol sequences. 
Here we have a new result. The boundary of yellow and orange 
regions corresponds to the binary collision curve (broken curve) of 
type 3 (Tanikawa et al. 1995) emanating from point $T_2$ in Fig. 5. 
In Tanikawa et al. (1995), we could not follow the collision curve 
down to the $x$-axis due to numerical difficulty. In the present 
work, It has been shown that the collision curve actually arrive at 
the $x$-axis. 

\begin{figure}[h] 
\begin{center}
\plotfiddle{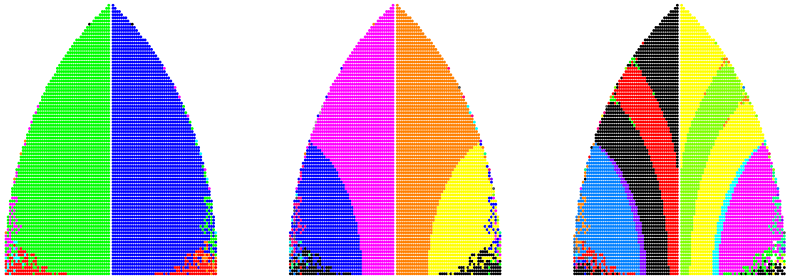}{3 cm}{0}{80}{80}{-200}{-60}
\end{center}
\caption{The structure of the initial condition plane. 
From the left, the first two digits of symbol sequences, 
the first three digits, and the first four digits.}
\end{figure}

In the right panel of Fig. 3, the results for the first four digits 
are shown. New collision curves appear. One is the boundary between 
light blue and pink regions. This is the collision curve of type 1 
emanating from $T_2$. The other two curves between yellow and green 
regions are a collision curve emanating from $T_{1,1}$ and a curve 
of type 2 starting at the middle of $T_1$ and $T_{1,1}$ in Fig. 5.

\subsection{The first digit for systems with angular momentum}

In the preceding secion, we have numerically shown that the boundaries 
of regions with different cylinders are formed with binary collision 
curves. In this section, let us look at the initial condition plane from 
a different view point. We gradually add angular momentum to triple 
systems. The parameter is the virial coefficient $k = T/|U|$ where 
$T$ is the kinetic energy and $U$ is the potential energy of a triple 
system. As $k$ increases from 0, The contibution of velocities increases
relative to a fixed potential energy. 
At $k=1$, the energy of the triple system 
is equal to zero. So we consider the range $0 \leq k \leq 1$. 

\begin{figure}[h] 
\begin{center}
\plotfiddle{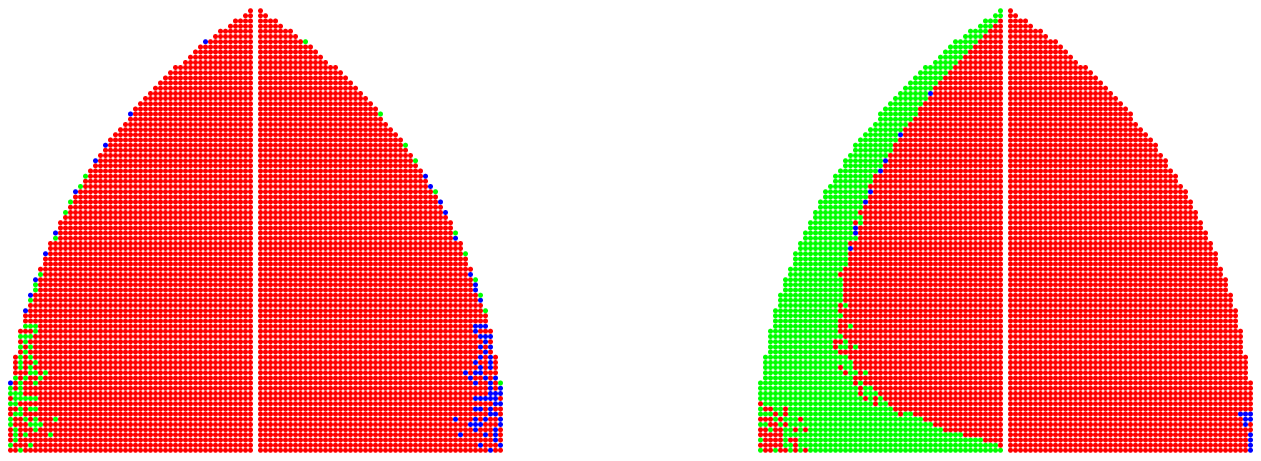}{1 cm}{0}{50}{50}{-120}{-60}
\plotfiddle{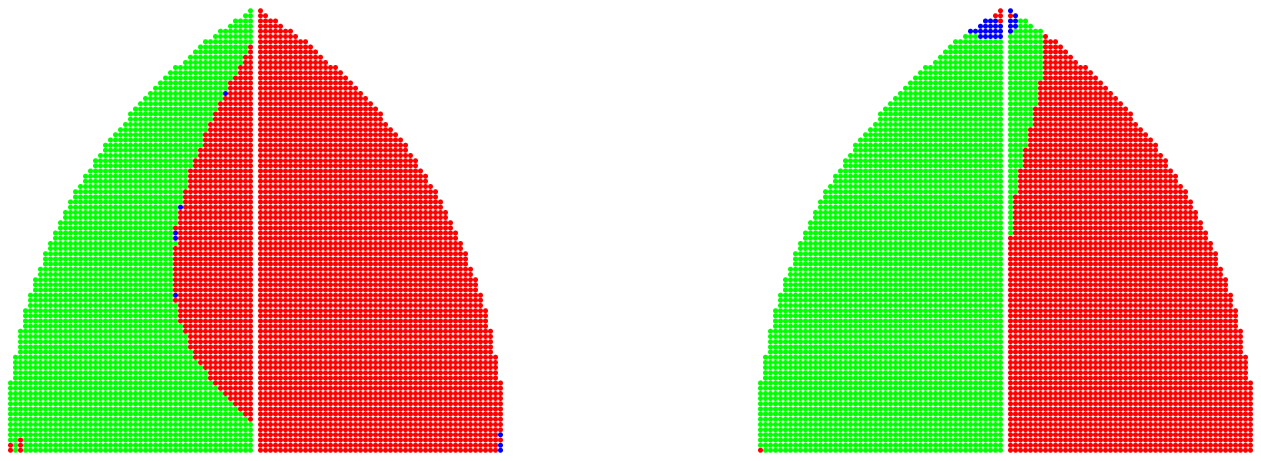}{3 cm}{0}{50}{50}{-120}{-40}
\plotfiddle{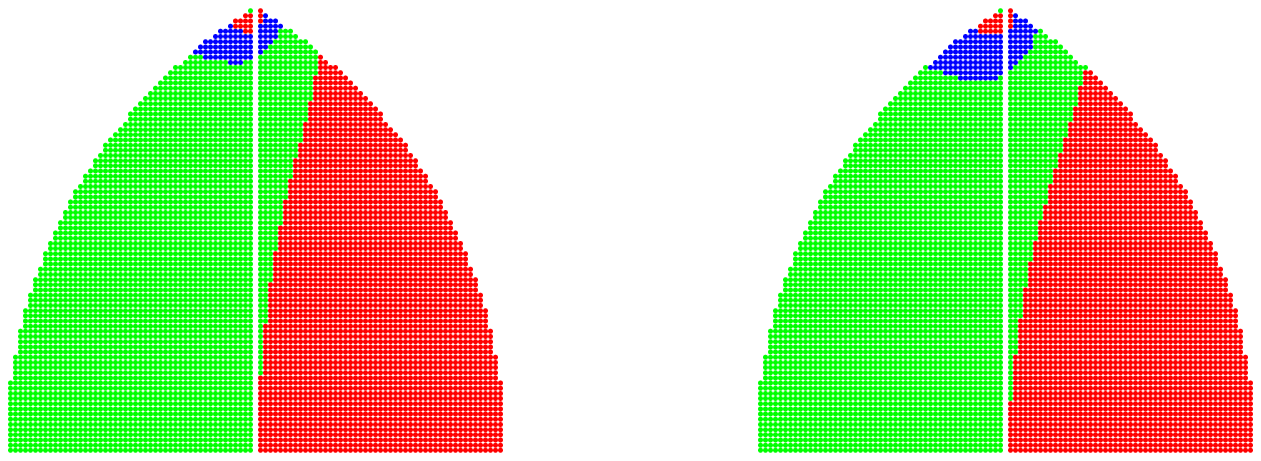}{1 cm}{0}{50}{50}{-120}{-75}
\plotfiddle{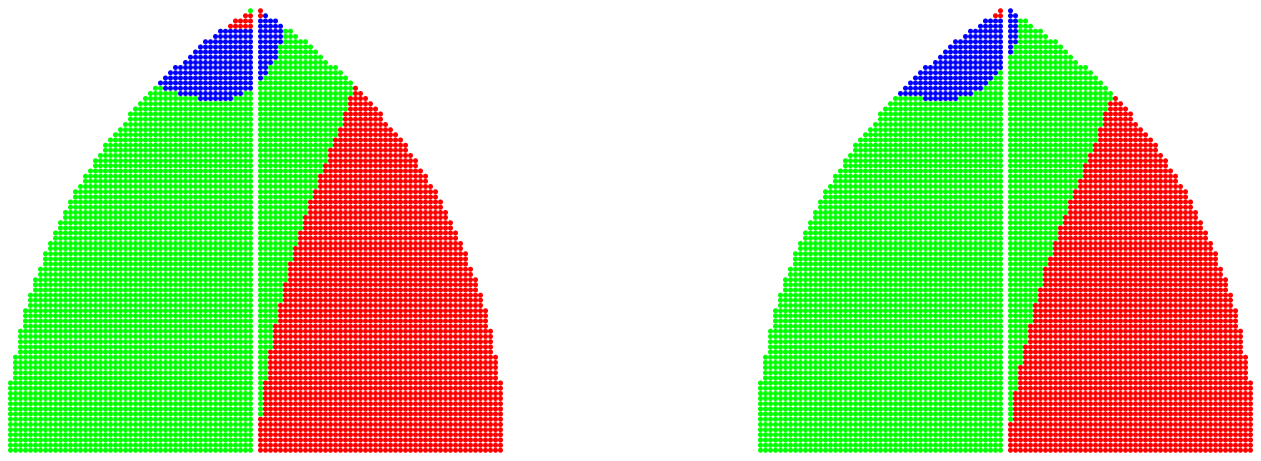}{3 cm}{0}{50}{50}{-120}{-50}
\end{center}
\caption{ The structure of the initial condition plane 
for $k=0, 0.001$ (top), $0.01, 0.1$ (second top), $0.2, 0.3$ (third 
top), and $0.5, 0.7$ (bottom) for the first digit of symbol sequences. 
Red is for symbol '1', green is for symbol '2', and blue is 
for symbol '3'.
}
\end{figure}

In this report, we consider only the boundaries of the cylinders of 
the first one word as a function of $k$. In Fig. 7, results are given 
for $k=0, 0.001$ (top), $0.01, 0.1$ (second top), $0.2, 0.3$ (third top), 
and $0.5, 0.7$ (bottom). 
Red is for symbol '1', green is for symbol '2', and blue is 
for symbol '3'.

As mentioned before, the whole plane is occupied by region '1' 
for $k=0$. Even for a very small $k>0$, region '2' appears 
(top-right figure). This is because near the left boundary of the plane 
bodies 1 and 2 form a binary, and even for a small angular momentum, 
this binary revolves each other, hence body 2 instead of body 1 
experiences the first syzygy crossing. The boundary of two regions is 
formed with a binary collision curve of type 3 (collision bewteen bodies 
1 and 2). It seems that this curve extends from the center bottom (the 
Euler point) to the top (the Lagrange point). 

As $k$ increases, the boundary collision curve deforms and 'evolves' 
to the right in the initial condition space (see the second top 
figures of Fig. 7). Near the top of the plane, region '3' appears. 
Though this region is visible only for $k \geq 0.1$ in Fig. 7, 
the region is visible for $k = 0.001$ if we enlarge the neighborhood of 
the Lagrange point. In addition, the binary collision curve seems to 
spiral into the Lagrange point as long as the point is unstable 
(see Fig. 8). 

\begin{figure}[h] 
\begin{center}
\plotfiddle{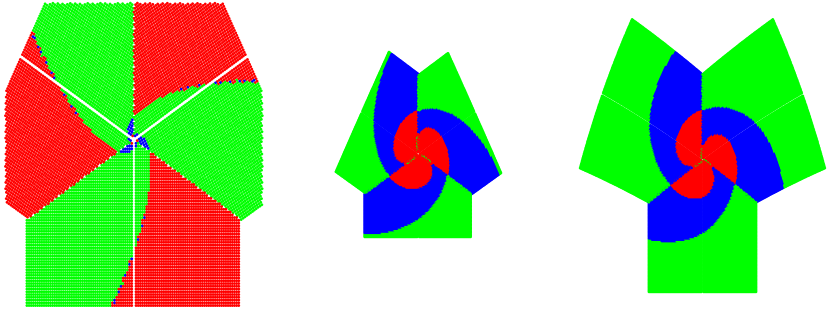}{3 cm}{0}{80}{80}{-200}{-60}
\end{center}
\caption{The structure of the initial condition plane around the 
Lagrange point for the first one digit. 
$k = 0.001$ (left) for $-0.005 \leq x \leq 0.005$, 
$0.86 \leq y \leq 0.87$; $k=0.01$ (center) for 
$-0.01 \leq x \leq 0.01$, $0.85 \leq y \leq 0.87$, and $k=0.1$ (right)
for $-0.1 \leq x \leq 0.1$, $0.76 \leq y \leq 0.96$.}
\end{figure}

\vspace{0.3cm}
\noindent
\section{Conclusions}

We have demonstrated that symbolic dynamics is effective in the 
planar three-body problem. Main results are

\vspace{0.3cm}
\noindent
(i) In the free-fall problem, it has been shown that binary collision
curves divide the initial condition space. This has been suggested in 
Tanikawa et al. (1995). In the present work, divisions are more clearly 
seen. 

\vspace{0.3cm}
\noindent
(ii) It has been shown for the three-body problem with angular momentum 
that a binary collision curve connects the Euler point and 
Lagrange point in the initial condition plane, which suggests 
that the Euler and Lagrange points are connectd by an invarian manifold.  
The collision curve seems to spiral into the Lagrange point if the 
angular momentum is not zero. 

\vspace{0.3cm}
\noindent
{\bf Acknowledgment}

This project was sipported by JSPS and RFBR under the Japan--Russia
Research Coorperative Program. 
One of the authors (K.T.) expresses hearty thanks to the Russian 
members of Japan-Russia Joint Research for the hospitality during  
his stay in Sgt Petersberg in the end of August, 2007.

\end{document}